# THE NUMBER AND OBSERVABILITY OF POPULATION III SUPERNOVAE AT HIGH REDSHIFTS

SIMONE M. WEINMANN
SIMON J. LILLY

Department of Physics, ETH-Zurich, CH-8093 Zurich, Switzerland

## ABSTRACT

We consider the feasibility of detecting Population III pair-instability supernovae (PISN) at very high redshifts with the James Webb Space Telescope (JWST). Four published estimates for the PISNe rate show a rather wide dispersion, between 50-2200 deg$^{-2}$ yr$^{-1}$. Correcting problems with several of these, we conclude that even a fairly optimistic estimate is probably a further order of magnitude lower than this range, at a rate of order 4 deg$^{-2}$ yr$^{-1}$ at $z \sim 15$ and 0.2 deg$^{-2}$ yr$^{-1}$ at $z \sim 25$, both with substantial uncertainty. Although such supernovae would be bright enough to be readily detectable with the JWST at any relevant redshift, the lower number densities derived here will likely require either a dedicated wide-angle search strategy or a serendipitous search. We expect that typically about 1 deg$^2$ (or 500 JWST NIRCam images) per detected supernova at 4.5 μm must be imaged to detect one PISN at $z \sim 15$ and about 35 deg$^2$ to detect one at $z \sim 25$. If some Population III star-formation persists to lower redshifts $z \sim 5$, then PISNe may also be detectable in wide-angle ground-based Z-band imaging surveys at $Z_{AB} \sim 23$, at a density of order 1 deg$^{-2}$ of surveyed area. In an Appendix, we consider the possible effects of intergalactic dust in obscuring high redshift supernovae or other high redshift sources. We show that the obscuration at a given rest-wavelength will peak at some maximum redshift and thereafter decline. While it may be a significant effect in observations of the very high redshift Universe, it is unlikely, even under rather pessimistic assumptions, to completely obscure primordial objects.

# 1. INTRODUCTION

The first stars to form in the Universe are predicted to be very massive (e.g. several hundred $M_\odot$, e.g. Bromm & Larson 2003), very metal poor and short-lived. Depending on their mass, these massive stars should end their lives either by collapsing into black holes or by being completely disrupted in violent pair-instability supernovae, PISNe (e.g. Fryer et al 2001, Heger & Woosley 2002). PISN explosions are expected to occur for stars with a relatively narrow range of masses between $140 < M/M_\odot < 250$ - see Heger et al (2002) for a review.

Pop III stars are predicted to first appear in the Universe at some point between $30 < z < 50$ (see e.g. Yoshida et al 2003, Cen 2003, Mackey et al 2003, Wise and Abel 2003), and to cease being formed when the metallicity exceeds a critical metallicity $Z_{crit}$ that is estimated to be in the range $Z_{crit} \sim 10^{-4}$-$10^{-3}$ $Z_\odot$ (see e.g. Oh et al 2001, Bromm and Larson 2003). This threshold may well be exceeded as early as $z \sim 15$ (Mackey et al 2003, Yoshida et al 2003) but some Pop III star-formation could conceivably survive alongside higher metallicity star-formation to redshifts $z \sim 5$ if the mixing of metals in the intergalactic medium is incomplete (see e.g. Scannapieco et al 2003 for detailed models).

The formation of the first stars in the Universe, their initial mass function and the chemical enrichment history of the intergalactic medium, as well as the detailed physics of PISN themselves, are all highly uncertain theoretically and will likely be guided by future observations by facilities such as the James Webb Space Telescope (JWST).

The putative PISNe that may be associated with Pop III would, at maximum, be of order $10^5$ brighter than the Pop III stars themselves, and detection of these transients may be the best hope for detecting Pop III at very high redshifts with future observing facilities such as the JWST, especially if the first stars are formed singly in individual dark matter haloes (Abel et al 2002). The detection of such massive explosions in the early Universe would be of great interest in determining the physical conditions in and around the first structures to form in the Universe, including quantifying feed-back loops operating to control the star-formation in young haloes, and in understanding the chemical enrichment and reionization of the Universe.

It should be noted that the existence and importance of PISN in the early chemical and reionization history of the Universe is not beyond question. For instance, Tumlinson et al (2004) have argued against a large role for PISN in Pop III on account of the abundance patterns of low metallicity Galactic Pop II stars.

There have been several calculations of the expected rate of such transients often as a side product of investigations primarily aimed at other aspects of star and structure formation in the early Universe. Unfortunately, these published PISNe rate predictions show a rather wide variation, between 50-2200 deg$^{-2}$yr$^{-1}$ (see Mackey et al 2003, Heger et al 2002, Cen 2003, Wise & Abel 2003). In this paper, we examine the causes for this variation and conclude that the most likely (but still ultimately optimistic) rate is substantially lower even than the lower end of these published estimates. This leads us to consider the optimum strategy for searching for PISN at very high redshifts with the JWST.

Then in the last section, we consider the possibility of observing Pop III PISNe at low redshifts. We consider Z-band photometric searches at $z \sim 5$ and report on the results of a preliminary but inconclusive search at this redshift.

In an Appendix, we examine the possible effects of intergalactic dust in obscuring high redshift supernovae or other high redshift sources, and show that the obscuration at a given rest-wavelength peaks will peak at some maximum redshift and thereafter decline. While it may represent a significant effect, intergalactic dust obscuration is unlikely to completely obscure very high redshift objects.

Throughout the paper we assume a concordance cosmology with $H_0 = 70$ kms$^{-1}$Mpc$^{-1}$, $\Omega_m = 0.25$, $\Omega_b = 0.04$ and $\Omega_\Lambda = 0.75$.

## 2. PUBLISHED PISNe RATES

There have been, as far as we are aware, four independent calculations of the Pop III supernovae rates at very high redshifts (see also the calculation by Miralda-Escudé & Rees 1997 and Madau et al 1998a at lower redshifts). Unfortunately, given the potential importance of these as observational tracers of Pop III, they display a broad range of between 50-2200 deg$^{-2}$yr$^{-1}$ (Mackey et al 2003, Heger et al 2002, Cen 2003, Wise & Abel 2003). As well as variations in input assumptions, three of these calculations appear to have additional problems. Accordingly, we here review the appropriate calculation.

Neglecting the short lifetime of PISN progenitors and assuming a narrow mass range of PISN progenitors, the comoving PISNe rate in terms of the star-formation rate (SFR) is:

(1) $$\frac{dN}{dt} = SFR(z) \times \mathcal{N}$$

$\mathcal{N}$ is the specific supernova number, i.e. the number of supernovae produced per unit stellar mass created, which will be approximated by the mass fraction (within the initial mass function) that is in supernovae producing stars divided by $m_{SN}$, the mass of each PISN progenitor.

(2) $$\mathcal{N} = \frac{\int_{m_1}^{m_2}\left[\frac{dN}{d\log m}\right]m^{-1}dm}{\int_0^{\infty}\left[\frac{dN}{d\log m}\right]dm} \approx \frac{f_{SN}}{m_{SN}}$$

The *SFR(z)* may be written in terms of the fraction of baryonic material in the Universe that is cumulatively formed into Pop III stars, $f_{III}$, as

(3) $$SFR(z) = \Omega_b \rho_c \times \frac{df_{III}}{dt}$$

Thus the observed PISNe rate per unit solid angle on the sky is

(4) $$\frac{dN}{dtdz} = SFR(z) \times \mathcal{N} \times \left(D^2 \frac{dD}{dz}\right) \times \frac{1}{1+z}$$

where the second term in brackets is the comoving volume element in a spatially flat Universe, *D* being the comoving distance. The final (1+z) factor accounts for the cosmic time-dilation of the rate as observed by us. Together with (3), this simplifies to the following since $dD/dt = c(1+z)$:

(5) $$\frac{dN}{dtdz} = -(\Omega_b \rho_c) \times \mathcal{N} \times D^2 \times c \times \frac{df_{III}}{dz}$$

If the star-formation rate as expected decreases strongly with redshift, then, since *D* is almost constant at very high redshifts,

(6) $$\frac{dN}{dt}(z > z') \sim \Omega_b \rho_c \, \mathcal{N} D^2 c \, f_{III}(z')$$

The simple form of equations (5) and (6) shows that the observed supernova rate depends primarily on two quantities: the build-up of stellar mass $f_{III}$, and most critically the specific supernova number $\mathcal{N} \sim f_{SN} m_{SN}^{-1}$. The redshift and cosmology enter only rather weakly through the *D(z)* term.

Comparing with equation (6), the Mackey et al (2003) calculation is clearly correct. The Heger et al (2002) rate, which has been used also by Panagia (2003), is high by $(1+z)^2$. We think also that Cen's (2003) estimate is a factor of (1+z) high (there should not be a (1+z) term in equation 24) and may also have a small numerical problem in the computation of $N_{HN}$. Finally, Wise & Abel's (2003) estimate, based on placing one PISN in each newly formed dark matter halo contained a numerical error (Abel and Wise, private communication). These different estimates are summarized in Table 1, where the final column lists the "adjusted" rates correcting for the above problems. Once adjusted, the resultant supernovae rates broadly scale in the expected way with the input parameters $f_{III}$ and the $f_{SN}m_{SN}^{-1}$ product.

## 3. ADOPTED RATE AND JWST SURVEY STRATEGY

The correct calculation, by Mackey et al (2003), has in our opinion (see below) relatively optimistic input assumptions, e.g. $f_{SN} \sim 1$ and $m_{SN} \sim 250$ M$_\odot$ and thus $\mathcal{N} \sim 0.004$, thereby masking the differences between these different published estimates.

For the $f_{III}(z)$ parameter, we consider the range of star-formation rates presented by Yoshida et al (2003), Ricotti et al (2003), Mackey et al (2003) (omitting star-formation at $z > 30$), Somerville et al (2003) and Scannapieco et al (2003). These are shown in Fig 1. They generally all exhibit the roughly exponential behaviour, SFR $\propto \exp(-z/3)$ that is a rather generic prediction of high redshift star-formation that is dominated by the proliferation of dark matter haloes in the exponential part of the Press-Schechter mass function (see Hernquist & Springel 2003). A reasonable summary of the available models is that at redshift $z \sim 15$, $\log f_{III} \sim -3.7$, with a range of $\pm 0.5$ dex and by $z \sim 25$ it has likely fallen to $\log f_{III} \sim -5 \pm 1$. In what follows we take a Pop III star-formation history given by the solid line in Fig 1.

The value of $\mathcal{N}$ is even less certain. Even assuming that the physics of PISN progenitors is well understood, it is strongly dependent on the initial mass function. Following Nakamura and Umemura (2001) we will assume that the initial mass function is bimodal, with one component containing a fraction $\kappa$ of the integrated stellar mass with an initial mass function peaking at low masses (of order 1 M$_\odot$) and a second component peaking at much higher masses and containing (1-$\kappa$) of the total mass. If we assume that this second high mass component is represented by a power-law between mass cut-offs at $M_1$ and $M_2$.

$$(7) \quad \frac{dN}{d \log M} = M^{-\alpha} \quad \text{for } M \geq M_1$$
$$= 0 \quad \text{for } M < M_1 \text{ and } M > M_2$$

and that PISN are produced between $M_{low}$ and $M_{high}$ (assumed to be greater than $M_1$) then $\mathcal{N}$ will be:

$$(8) \quad \mathcal{N} = \frac{1}{M_1} \times (1-\kappa)\left(\frac{\alpha-1}{\alpha}\right)\left(\frac{M_1}{M_{low}}\right)^\alpha \left\{1 - \left(\frac{M_{high}}{M_{low}}\right)^{-\alpha}\right\}\left\{1 - \left(\frac{M_2}{M_1}\right)^{1-\alpha}\right\}^{-1}$$

We tabulate some representative values of $\mathcal{N}/(1-\kappa)$ in Table 2 for various choices of $\alpha$ (i.e. $\alpha = 1.3$ and $\alpha = 3$, following Nakamura and Umemura 2001) and $25 < M_1 < 100$ M$_\odot$, assuming $M_{low} = 140$ M$_\odot$ and $M_{high} = 250$ M$_\odot$ and truncating the i.m.f. at $M_2 = 1000$ M$_\odot$.

Given that there is likely to be a strong mass-luminosity relation for PISNe (see e.g. Wise & Abel 2003) it is likely that the lowest mass PISNe may not be observable and thus we may modify $\mathcal{N}$ to exclude those PISNe by modifying the lower mass cut-off in (8) so that $M_{low} = M_{vis}$, the mass of the least massive detectable PISN, to yield $\mathcal{N}_{vis}$. Table 2 lists some values of $\mathcal{N}_{vis}$ for $M_{vis} = 200 M_\odot$ corresponding to a 1.7 magnitude drop in peak luminosity relative to 250 $M_\odot$ progenitor.

An absolute upper limit is obviously $\mathcal{N} \sim M_{low}^{-1} \sim 0.007\ M_\odot^{-1}$, but this would require all stars to have been formed with a single optimal mass, $M = M_{low}$. Mackey et al (2003) take $\mathcal{N}$ to be only slightly lower than this extreme value $\mathcal{N} \sim 0.004\ M_\odot^{-1}$. Inspection of the values of $\mathcal{N}_{vis}$ in Table 2 suggests that for practical surveys, values of $\mathcal{N}_{vis} \sim$ few $\times 10^{-4}\ M_\odot^{-1}$ arise from quite plausible i.m.f.'s, e.g. $\alpha = 1.35$ and $\kappa = 0.5$. We have somewhat arbitrarily adopted a value of $\mathcal{N}_{vis} \sim 4.4 \times 10^{-4}\ M_\odot^{-1}$ with an uncertainty of 1 dex, which then dominates over the uncertainty in $f_{III}$. This is an order of magnitude lower than Mackey et al's $\mathcal{N}$ accounting for the difference in observed supernova rates.

With the above assumptions, these translate to PISN rates of between $0.3 - 30$ PISN yr$^{-1}$deg$^{-2}$ at $z > 15$ and between $0.01 - 3$ yr$^{-1}$deg$^{-2}$ at $z > 25$, i.e. substantially below the original estimates that have appeared in the literature. It should be noted that the difference with Mackey et al (2003) is almost entirely due to the lower assumed value of $\mathcal{N}$ or $\mathcal{N}_{vis}$. The differences with the other, even higher estimates discussed above reflect the additional problems identified earlier.

Finally, it should be noted that these supernova rates are three orders of magnitude lower than those calculated for SN II at lower redshifts by Miralda-Escudé & Rees (1997) in the context of the enrichment and reionization of the intergalactic medium at $z \sim 5$. The difference is because the above assumptions lead to one supernova per $5 \times 10^6\ M_\odot$ of baryonic material at $z \sim 15$ compared with the one per 5000 $M_\odot$ taken by them at $z \sim 5$. This is due to both the larger mass of stars formed by $z \sim 5$ and the larger $\mathcal{N}$ associated with lower mass Type II SNe.

The number of supernovae that are present in a particular image of an area of sky is given by

$$(9) \qquad N \sim \frac{dN}{dt} A\, \Delta t_{vis,0} (1+z)$$

where $A$ is the solid angle area of the image and $\Delta t_{vis,0}$ is the time interval in the rest-frame during which the PISN would remain above the brightness detection threshold of the image. The $(1+z)$ factor accounts for the time-dilation of the observability period in the observed frame. It is this surface density of detectable transients that determines the survey efficiency in detecting PISNe, provided only that the sampling strategy allows repeat visits after an interval $\Delta t_{vis}$ to optimally detect transients.

Based on PISN models (Heger et al 2002), the maximum brightness of a PISN near the upper end of the progenitor mass range (250 $M_\odot$) is roughly independent of wavelength across a broad range of wavelengths longward of Lyα (out to at least rest-frame 0.5μm) and is roughly AB ~ 26 (see Fig 2) for $15 < z < 25$. In detail, the maximum brightness is dimmer by about 0.5 magnitudes at the longer wavelengths (Heger et al 2002). Detection of sources at AB ~ 27, i.e. of order one magnitude below maximum, is essentially impossible from the ground in the near-infrared, but should be straightforward with the JWST NIRCam (Rieke et al 2003) and, scaling from today's 6-10m telescopes, would be feasible with a ground-based 30-m telescope shortward of 2.5μm. A point source with AB ~ 27 requires a nominal integration time of less than 1000 seconds with the JWST NIRCam for a 10σ or greater detection over the 1.5-4.5 μm waveband (Rieke et al 2003). Thus, in terms of sensitivity, JWST should have no trouble detecting PISNe to $z \sim 50$, if they exist.

It is important to note that, in the Heger et al (2002) models, the duration of the time above a given brightness threshold, set relative to the maximum brightness, $\Delta t_{vis}$, increases quite strongly with wavelength (see Fig 3). At a threshold one magnitude below maximum, the rest-frame visibility period is only about 1.7 days at rest-1600 Å, 4.2 days at rest 2400 Å and 20 days at rest 5000 Å, i.e. increasing as roughly $\lambda^2$. The resulting increase in number of PISN per image at longer wavelengths implied by equation (7) more than compensates (at least for a detector with flat AB sensitivity) the roughly 0.5 magnitude decrease in peak brightness over this same wavelength interval.. Lowering the detection threshold further will also increase $\Delta t_{vis}$ but the gain in $\Delta t_{vis}$ goes as less than the square of the limiting flux (Fig 2), so the number of supernovae discovered is maximized by aiming to detect them within a magnitude or so of maximum, in as many fields as possible. Likewise, the mass-luminosity relation for PISN is apparently so steep (see e.g. Wise & Abel 2003) that it is better, in terms of simple number of supernovae to be detected, to aim for the brightest objects near the upper mass cut-off in the largest possible area of sky, rather than to try to probe further down the mass function in fewer fields.

The step in JWST sensitivity across 5 μm (going from NIRCam to MIRI) of about 2 magnitudes and the reduction of the field of view by a factor of 4 (Wright et al 2004) produces a decrease in survey efficiency of a factor of 25 for constant AB magnitude, meaning that the most attractive wavelength for PISN searches with JWST is probably at the long end of the NIRCam range, i.e. 3.5-4.5 μm.

We show in Fig 4 the number of PISN expected at different redshifts for deep imaging surveys carried out at different wavelengths below 5 μm, assuming that each survey reaches 1 magnitude below the brightness of maximum light, as computed for that redshift and observing wavelength. We assume a star-formation rate given by the heavy curve in Fig 2. Numbers for other star-formation histories or other choices of $\mathcal{N}_{vis}$ are straightforwardly calculable, while those for deeper surveys can be computed using the visibility curves in Fig 3.

At 3.5-4.5 μm, we would expect to have to observe an average of order 1 deg$^2$ to detect one PISN at $z > 15$ in a blind search reaching to roughly 27th magnitude, equivalent to 500 NIRCam images. This could be built up through serendipitous surveys, or, since the required exposure time is relatively short, under 1000s (Rieke et al 2003), through an optimized search program requiring of order $4\times10^5$ seconds per supernova. For maximum detection efficiency images of the same field would have to be separated by more than 90 days for $z \sim 15$. Only about 200 NIRCam images (0.4 deg$^2$) would be required if the depth was increased by 2 magnitudes, with an optimum separation between images of 8 months or more, but would require much longer exposures to gain the extra depth.

At higher redshifts, the reduction in $f_{III}$ (and $\Delta t_{vis,0}$ at fixed observed λ) leads to a rapid increase in area that must be surveyed, e.g. to of order 35 deg$^2$ PISN$^{-1}$ at $z > 25$, i.e. one per 16,000 images (with images separated by at least 50 days) for supernovae within one magnitude of maximum. We note that this estimate is probably larger than the total number of NIRCam images to be obtained in the entire JWST lifetime.

Thus, unless our ideas about the early Universe are completely wrong, it is likely that JWST's reach in redshift space through stand-alone supernova surveys will be limited to somewhere in the $15 < z < 25$ range by the rareness of the transient events, rather than by their faintness. "All-sky" or at least "wide-field" searches for transients at other wavelengths, such as Gamma Ray Bursts (GRB), followed by prompt follow-up with the JWST would extend the reach to higher redshifts, but only if PISN are associated with such events.

Restricting attention to wavelengths in the K-band (2.2 μm) as would be appropriate for a possible ground-based survey (e.g. with a thirty meter telescope or larger) we find that one would need, statistically, to image of order 4.5 deg$^2$ to detect one PISN at $z \sim 15$, with the two epochs separated by at least 20 days. This is a somewhat lower discovery rate than with the JWST because the shorter wavelengths decreases the visibility period $\Delta t_{vis,0}$. Intervening neutral Hydrogen absorption almost certainly kills PISNe at only slightly higher redshifts ($z > 17$). It should be noted that the proposed SNAP satellite (Aldering et al 2004) is not designed for wavelengths beyond the *H*-band (1.6 μm) and therefore is unlikely to detect sources at redshifts much above $z \sim 10$.

## 4. PISNe AT LOWER REDSHIFTS

Scannapieco et al. (2003) have suggested that Population III stars may still be forming down to a redshift as low as $z \sim 5$ in regions of the Universe that have not been polluted by metals due to inefficient mixing of the IGM. Such objects could be detected in deep wide-field imaging surveys in the *Z*-band (λ ~ 0.95 μm). Using one of their representative Pop III star-formation rates at $z = 5$, we derive a PISN rate of about 25 yr$^{-1}$ deg$^-$

$^2$ per unit redshift. The peak brightness of the PISNe light curve at $z = 5$ should be around $Z_{AB} = 23$, i.e. well within the detection limits of wide-field CCD imagers on 4-8m telescopes.

Expecting that the SNe would be visible for about 10 days in the observed frame (since only the region around Ly$\alpha$ is detectable) within one magnitude of maximum yields an estimated rate of one PISN per 0.7 per deg$^2$ of images, making it attractive to search for such systems in large scale imaging surveys that are now possible with instruments like the Megacam (Boulade et al 2003) on the 3.6m CFHT.

A sensitive wide-field search in the near-infrared to 24-25 magnitude, feasible with a suitably instrumented wide field 8-m telescope, would detect more $z \sim 5$ PISNe deg$^{-2}$ because of the increased $\Delta t_{vis,0}$.

## 5. SUMMARY

We have investigated the potential observability of PISNe from Pop III stars at $15 < z < 25$ with the JWST and other facilities, with the following conclusions:

1. We find that the expected PISNe rate is much lower than previous published estimates, probably around 4 deg$^{-2}$yr$^{-1}$ at $z \sim 15$ and 0.2 deg$^{-2}$yr$^{-1}$ at $z \sim 25$ rather than the 50-2500 deg$^{-2}$yr$^{-1}$ that has been quoted in the literature.
2. The brightness of such supernovae is such that they could be detected in a relatively short (under 1000s) exposure with NIRCam on JWST. It will be best to search at the longest practicable wavelengths to take advantage of the longer visibility time of the PISNe as well as to be sensitive to PISNe at the highest possible redshifts. For the JWST, the optimum wavelength is likely to be at about 3.5-4.5 μm using the NIRCam camera.
3. However, with our lower PISNe rate, the chances of detecting a PISN on any individual JWST NIRCam frame will be quite small and we would expect that about 1 deg$^2$ (i.e. 500 NIRCam images) would have to be taken to detect each PISN at $z \sim 15$ at one magnitude below maximum, or 0.4 deg$^2$ (200 images) three magnitudes below maximum. The former strategy is practical either through serendipitous searches or through a dedicated wide-field survey taking relatively short exposures.
4. At higher redshifts, the surface density of detectable supernovae should drop precipitously, requiring surveys of order 10-35 deg$^2$ (i.e. up to 16,000 NIRCam images) per supernova detected at $z \sim 25$. This probably exceeds what is possible in the JWST mission. Thus utilization of the full power of JWST to study transients at very high redshift may involve "all-sky" or very wide-field searches at other wavelengths (such as Gamma Ray Bursts) with JWST follow-up. It is of course not clear if PISN would be associated with GRB.

6. Finally in an Appendix, we show that foreground intergalactic dust produced by Pop III PISNe is unlikely to seriously affect the detectability of sources even to the highest redshifts. It has the interesting property that the extinction for light at a given rest-wavelength first increases but then decreases at the highest redshifts as soon as the comoving density of dust starts to drop faster than $(1+z)^{-1.5}$.

Our analysis illustrates a general problem with searches for the signatures of "First Light" in the Universe. Given that the Pop III star-formation rate is likely to decline exponentially with redshift at the highest redshifts, one will inevitably be searching for rarer and rarer objects on the sky. We have seen above that a search with JWST at $z \sim 15$ will likely succeed, but that the JWST is unlikely to discover PISN at much higher redshifts (e.g. $z \sim 25$) since it will simply not observe a large enough area of sky (of order 30 deg$^2$ required) even in the mission lifetime.

We thank Cristiano Porciani, Tom Abel, and Marcella Carollo for useful discussions and David Schade at HIA for his prompt assistance in identifying suitable $Z$-band images in the CFHT Archive. SJL is an ESA-supported Interdisciplinary Scientist for JWST.

**TABLE 1**

|  | redshift | $f_{III}$ | $m_{SN}$ | $f_{SN}$ | $\mathcal{N}$ | rate yr$^{-1}$ deg$^{-2}$ | |
|---|---|---|---|---|---|---|---|
|  |  |  |  |  |  | original | adjusted |
| Mackey et al | $z > 15$ | $2.7\times10^{-4}$ | 250 | 1 | $4\times10^{-3}$ | 50 | 50 |
| Heger et al | $z \sim 20$ | $10^{-6}$ | 250 | 1 | $4\times10^{-3}$ | 120 | 0.3 |
| Cen | $z > 13$ | $10^{-4}$ | 100 | 0.25 | $2.5\times10^{-3}$ | 580 | 11 |
| Wise & Abel | $z \sim 20$ | one per halo | | | | 2500 | |
| adopted here | $z > 15$ | $2\times10^{-4}$ | 225 | 0.1 | $4.4\times10^{-4}$ | $\sim 4$ | |
|  | $z > 25$ | $10^{-5}$ | 225 | 0.1 | $4.4\times10^{-4}$ | $\sim 0.2$ | |

**TABLE 2**

|  | $\mathcal{N}/(1-\kappa)$ (M$_\odot^{-1}$) | | $\mathcal{N}_{vis}/(1-\kappa)$ (M$_\odot^{-1}$)$^a$ | |
|---|---|---|---|---|
|  | $\alpha = 1.35$ | $\alpha = 3$ | $\alpha = 1.35$ | $\alpha = 3$ |
| $M_1 = 30$ M$_\odot$ | $8.4\times10^{-4}$ | $1.8\times10^{-4}$ | $2.5\times10^{-4}$ | $3.6\times10^{-5}$ |
| $M_1 = 50$ M$_\odot$ | $1.1\times10^{-3}$ | $5.0\times10^{-4}$ | $3.2\times10^{-4}$ | $1.0\times10^{-4}$ |
| $M_1 = 100$ M$_\odot$ | $1.6\times10^{-3}$ | $2.0\times10^{-3}$ | $4.7\times10^{-4}$ | $4.1\times10^{-4}$ |

Notes to Table 2:

(a) Taking $M_{vis}=200$M$_\odot$

# APPENDIX: POSSIBLE EFFECTS OF INTERGALACTIC DUST AT VERY HIGH REDSHIFTS

Population III PISN can pollute the intergalactic medium (IGM) effectively because their metal yield is high (Heger & Woosley 2002) and because supernova ejecta should be able to easily escape the shallow potential wells where such stars will be formed. Nozawa et al (2003) and Schneider et al (2004) suggest that as much of 15-30% of the progenitor star's mass may be transformed into dust grains in a PISN explosion. Since optical depth increases strongly with redshift for a species of constant comoving density, this poses the interesting question as to whether small amounts of intergalactic dust could obscure the highest redshift sources.

Elfgren & Desert (2003) have considered the evolution of the comoving dust density, $\Omega_d(z)$, at high redshift under several scenarios. The evolution is affected by many poorly constrained parameters, including the highly uncertain dust survival lifetime. As a toy-model, we assume their most pessimistic case which produces a dust density that rises with epoch to a value $\Omega_d \sim 10^{-5}$ to $z \sim 12$ and remains roughly constant thereafter to lower redshifts (see also Aguirre 1999b).

We adopt an LMC extinction curve $A(\lambda)$ from Weingartner & Draine (2001) but artificially set $A(\lambda) = \infty$ below Ly$\alpha$ to simply account for intergalactic HI absorption in the Gunn-Peterson trough. Assuming that the cosmological extinction is proportional to the comoving mass density of dust, i.e. that the size distribution of dust grains does not depend on redshift and is the same as in the LMC, the extinction for light observed at some wavelength $\lambda_o$ is given simply by:

$$(A1) \quad \frac{dA(\lambda_o)}{dz} = n_{H,0} \frac{dD}{dz}(1+z)^2 \times A(\lambda_o/1+z) \times \frac{[m_d/H]_z}{[m_d/H]_{LMC}}$$

where $A(\lambda)$ is the LMC extinction curve (taken from Fig 22 of Weingartner & Draine 2001), $n_{H,0} = 0.75\Omega_b\rho_c/m_H$ is the cosmic comoving hydrogen number density and $m_d/H$ is the dust mass per H atom, taken to be $[m_d/H]_z = \Omega_d/0.75\Omega_b$ and $[m_d/H]_{LMC} = 0.9 \times 10^{-27}$ gm (Weingartner & Draine 2001).

The extinction as a function of $\lambda_0$ for sources at different redshifts is shown in Fig 1 for this most pessimistic (in the sense of largest obscuration) Elfgren & Desert (2003) model for $\Omega_d$. As expected there is an appreciable increase in $A(\lambda_0)$ with increasing redshift for as long as the comoving dust density is roughly constant. However, at some point (see below), $\Omega_d$ will start to fall faster than $(1+z)^{-1.5}$, especially since the cosmic star-formation rate should have an exponential decline in $(1+z)$ (Hernquist & Springel 2003). Beyond this point, the extinction *for a given emitted wavelength*, e.g. Lyα, will actually decrease with increasing redshift and more distant sources will be less extinguished than foreground sources. This possibly counter-intuitive behavior is shown in Fig 1.

With this simple model, which we think is likely to overestimate the extinction, the absorption for emitted Lyα increases, as roughly $0.1z$ to $z \sim 12$, but then falls thereafter. We conclude that intergalactic extinction may be significant in modifying the brightness of very high redshift sources, but is unlikely to be so severe as to render them undetectable. At a fixed observing wavelength, the extinction of course increases monotonically with redshift. At a wavelength of 4.5 μm (see discussion below), it is predicted to be small.

It should also be noted that relative to the "average" Pop III star-formation history and characteristics that are assumed below (see Fig 2), we would need essentially all of the PISN progenitor mass to be converted to dust and efficiently distributed into the IGM to reproduce this $\Omega_d$. As noted above, a lower fraction by a factor of three or more is likely, suggesting that the results of this analysis should be regarded as an extreme worst case scenario, appropriate only if the actual Pop III star-formation rate is at the high end of those suggested in the literature. The extinction at $z \sim 1$ and $\lambda \sim 1$ μm in Fig 1 is about 0.03 magnitudes, equivalent to an undetectable rest-frame $E(B-V) \sim 0.01$ (see Riess et al 2000, Aguirre 1999a,b).

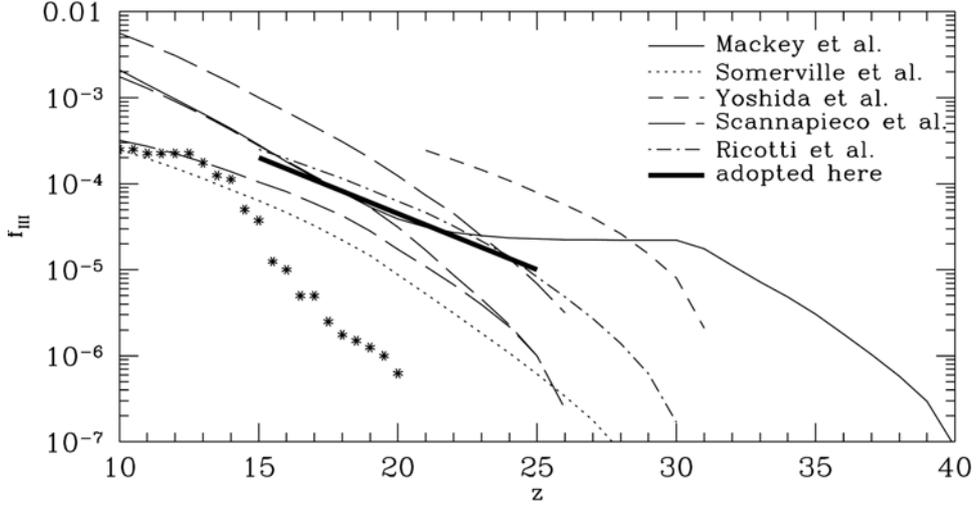

Fig 1. Cumulative production of Pop III stars from models in the literature, parameterised as $f_{III}$, the fraction of baryonic material that has been formed into Pop III. The heavy solid line shows the "average" production rate assumed in this paper, which must be considered uncertain by *at least* a factor of 3 at $z \sim 15$ and a factor of 10 at $z \sim 25$. The stars at lower left show the mass of dust assumed in the dust extinction model of Fig 1, which would be expected to be about $0.03 f_{III}$. It is therefore likely that the dust density used in Section 2 is overestimated by a factor of four or more (see Fig 1), unless the Pop III production is at the very high end of the models plotted here

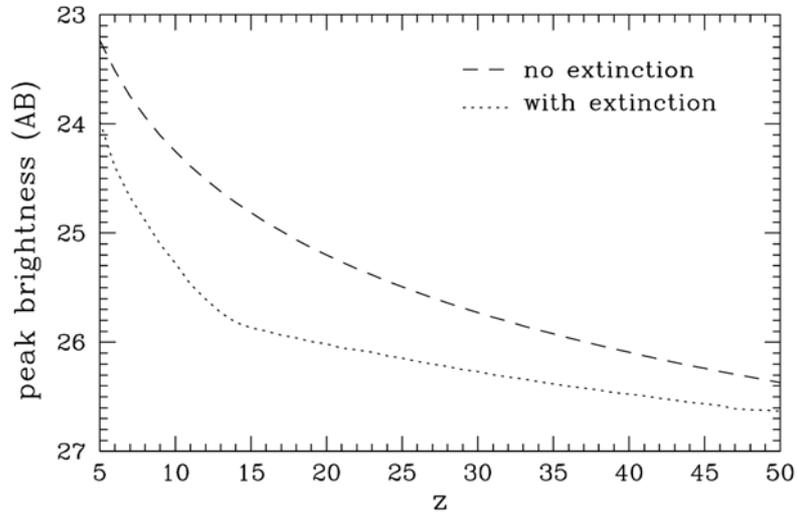

Fig 2. Observed peak brightness of 250 $M_\odot$ PISNe as a function of redshift in the spectral region around Lyα assuming both no significant extinction and the "worst case" extinction for the Lyα region shown in Fig A1 (adapted from Heger et al 2002). At wavelengths longward of Lyα, the peak brightness declines by only 0.5 magnitude to rest-frame 0.5μm wavelength. Note that these peak brightnesses are well within the detectability levels of JWST at all redshifts of interest. A 200 $M_\odot$ PISN is expected to be fainter by 1.7 magnitudes and a 175 $M_\odot$ one by 3.5 magnitudes.

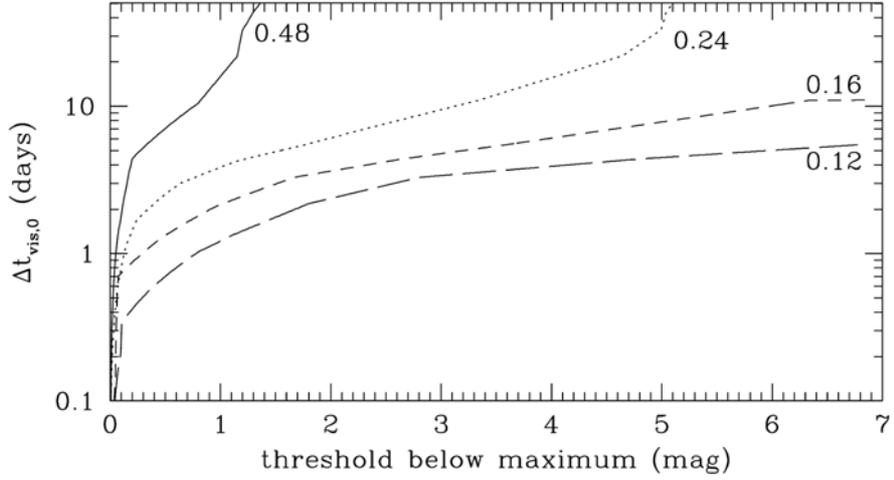

Fig 3. The variation of rest-frame visibility time $\Delta t_{vis}$ with the detection threshold (in magnitudes below maximum light) at different rest-wavelengths (adapted from Heger et al 2002). The increase in $\Delta t_{vis}$ with wavelength means that more PISNe are detected in a given area at longer wavelengths, more than outweighing their slightly decreased brightness. The optimum wavelength for searches with JWST is thus likely to be at the long end of the NIRCam spectral range, around 4-5 μm.

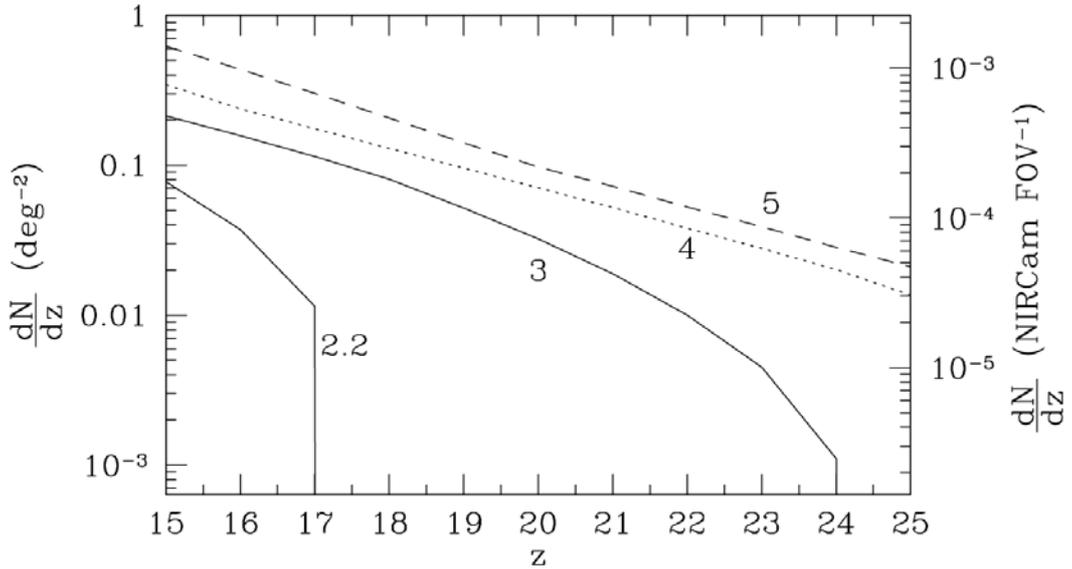

Fig 4. The surface density on PISNe on the sky per unit redshift as a function of redshift for four different observing wavelengths (labelled in μm), assuming that survey images reach one magnitude below maximum light (see Fig 3). These curves show the gain in the number of detected supernovae as the wavelength increases due to the increase in $\Delta t_{vis}$ in Fig 4 as well as the truncation in redshift due to disappearance of the supernovae below Lyα. The curves are normalised to $\mathcal{N}_{vis} = 4.4 \times 10^{-4}$ $M_\odot^{-1}$ as described in the text.

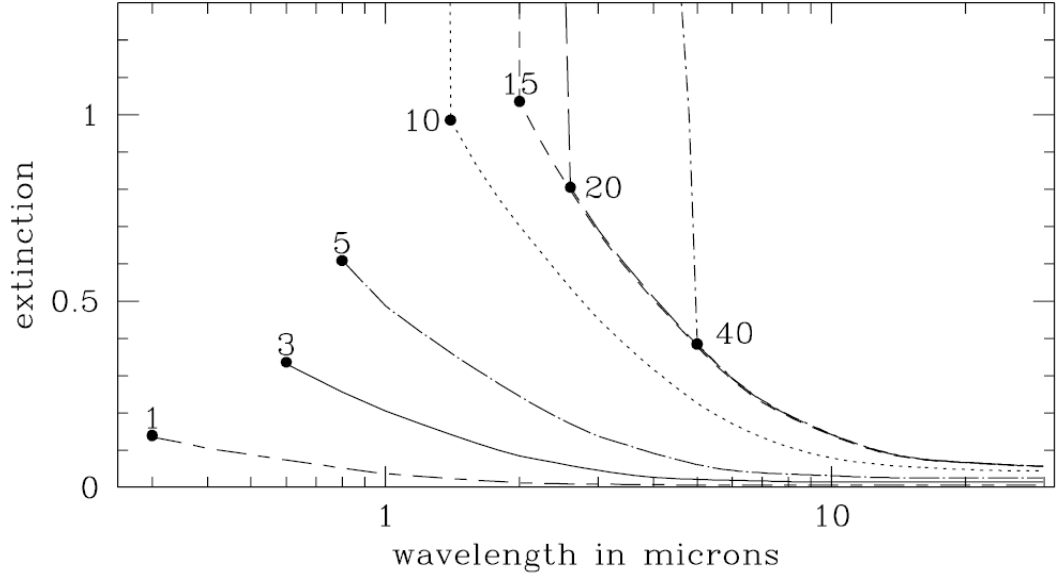

Fig A1. Extinction A(λ) in the observed frame for sources at increasing redshifts based on the most pessimistic dust model of Elfgren & Desert (2003). The solid points represent the observed wavelength of Lyα, labeled with the redshift of the extinction curve. Below Lyα the extinction is set to ∞ at all $z > 5$ to account for Gunn-Peterson HI absorption. It can been seen that for a given *rest-frame* wavelength (e.g. the dots representing Lyα) the extinction increases with redshift but then declines to higher redshifts as soon as the comoving dust density starts to fall faster than $(1+z)^{-1.5}$. At a given observed wavelength, the extinction must increase monotonically with redshift. It should be noted that the "average" Pop III star-formation rate adopted in this paper (see Fig 2) would produce at most only 25% as much dust as assumed here, so this curve can be considered as a very worst case. Even so, Lyα extinction is unlikely to prevent detection, and the extinction is likely to be very small at an observed wavelength of 4.5 μm (see text).